\documentclass[prl, twocolumn, final, reprint, amsmath, amssymb, superscriptaddress, showpacs, a4paper, aps, 10pt, hidelinks, numbers, sort&compress, floats, balancelastpage, longbibliography]{revtex4-1}

\usepackage{graphicx, color} 
\usepackage[normalem]{ulem}          

\usepackage[colorlinks=true]{hyperref}
\hypersetup{pdfpagemode=None, linkcolor=black, citecolor=blue, urlcolor=blue}

\usepackage[utf8]{inputenc}

\usepackage[caption=false]{subfig}
\usepackage{amssymb}
\usepackage{amsmath}
\usepackage{commath}
\usepackage{graphicx,bm}
\usepackage{verbatim}

\definecolor{britishracinggreen}{rgb}{0.0, 0.26, 0.15}
\definecolor{bulgarianrose}{rgb}{0.28, 0.02, 0.03}
\definecolor{darkred}{rgb}{0.90,0,0}
\definecolor{darkgreen}{rgb}{0,0.60,.2}
\definecolor{darkblue}{rgb}{0,0,1}
\definecolor{orange}{cmyk}{0,0.6,0.8,0}
\definecolor{lightblue}{rgb}{0.3,0.5,1}
\definecolor{lightgreen}{rgb}{0.4,0.80,.4}

\newcommand{\ssec}[1]{\noindent\emph{#1}\,---}

\newcommand{\Effg}{E_{\textrm{ffg}}}
\newcommand{\vF}{v_{F}}

\newcommand{\kF}{k_{F}}
\newcommand{\eF}{\varepsilon_{F}}
\newcommand{\vpb}{v_{\textrm{pb}}}
\newcommand{\Ntot}{N_\mathrm{tot}}
\newcommand{\vmax}{v_\mathrm{max}}

\begin{document}

\title{Dissipation Mechanisms in Fermionic Josephson Junction}

\author{Gabriel Wlaz\l{}owski}\email{gabriel.wlazlowski@pw.edu.pl}
\affiliation{Faculty of Physics, Warsaw University of Technology, Ulica Koszykowa 75, 00-662 Warsaw, Poland}
\affiliation{Department of Physics, University of Washington, Seattle, Washington 98195--1560, USA}

\author{Klejdja Xhani}\email{xhani@lens.unifi.it}
\affiliation{Istituto Nazionale di Ottica del Consiglio Nazionale delle Ricerche (CNR-INO), 50019 Sesto Fiorentino, Italy}

\author{Marek Tylutki}\email{marek.tylutki@pw.edu.pl}
\affiliation{Faculty of Physics, Warsaw University of Technology, Ulica Koszykowa 75, 00-662 Warsaw, Poland}

\author{Nikolaos P.~Proukakis}\email{nikolaos.proukakis@ncl.ac.uk}
\affiliation{Joint Quantum Centre (JQC) Durham-Newcastle, School of Mathematics, Statistics and Physics, Newcastle University, Newcastle upon Tyne NE1 7RU, UK}

\author{Piotr Magierski}\email{piotr.magierski@pw.edu.pl}
\affiliation{Faculty of Physics, Warsaw University of Technology, Ulica Koszykowa 75, 00-662 Warsaw, Poland}
\affiliation{Department of Physics, University of Washington, Seattle, Washington 98195--1560, USA}

\date{\today}

\begin{abstract}
We characterize numerically
the dominant dynamical regimes in a superfluid ultracold fermionic Josephson junction.
Beyond the coherent Josephson plasma regime, we discuss the onset and physical mechanism of dissipation due to the superflow exceeding a characteristic speed,
and provide clear evidence distinguishing its physical mechanism across the weakly- and strongly-interacting limits, despite qualitative dynamics of global characteristics being only weakly sensitive to the operating dissipative mechanism. Specifically,
dissipation in the strongly interacting regime occurs through the phase-slippage process, caused by the emission and propagation of
quantum vortices,
and sound waves –- similar to the Bose-Einstein condensation  limit. Instead, in the weak interaction limit, the main  dissipative channel arises through the pair-breaking mechanism.
\end{abstract}

\maketitle

\ssec{Introduction.} The Josephson effect of a supercurrent tunneling through a weak barrier has been one of the hallmarks of superfluidity. Originally it was proposed and realized as a junction between two superconductors separated by a thin insulating barrier~\cite{JOSEPHSON1962251,RevModPhys.36.216}.
With the advent of experiments with ultracold quantum gases, one can
produce a Josephson junction setup by splitting the atomic cloud into two parts through a relatively thin external potential~\cite{PhysRevLett.95.010402,PhysRevLett.111.205301,PhysRevLett.106.025302,Levy2007}. Controlling the number of atoms in both reservoirs, one
can produce either DC or AC Josephson junction. In the latter, the difference in chemical
potentials between both clouds plays the role of DC voltage used in the electronic Josephson junctions.
In such atomic systems, the tunability of interparticle interactions provides the means to controllably investigate in a single system
the dynamics, and stability, of such supercurrents across the BEC-BCS crossover~\cite{Valtolina2015,PhysRevLett.120.025302,PhysRevLett.124.045301,Kwon2020,Luick_2020},
including the strongly-interacting unitary Fermi gas (UFG) regime, which combines properties of the two limits.
Although one would expect different physical mechanisms to be at play in the limiting cases due to distinct low-lying excitation modes, understanding the dissipative dynamics in a fermionic Josephson junction across such regimes from a microscopic level remains an open question.

In this Letter, we provide clear evidence distinguishing the different physical mechanisms at play during  the dynamical evolution across the junction in a testable environment, thus extending beyond  the well-studied BEC regime~\cite{Levy2007,PhysRevLett.95.010402,PhysRevLett.106.025302,Smerzi03,PhysRevLett.79.4950,ZAPATA1998,PhysRevLett.124.045301,Xhani2020,KXfiniteT} for which simulations are relatively easy at the mean-field (Gross-Pitaevskii) level~\cite{RevModPhys.71.463}. Specifically, our numerical study -- which features no adjustable parameters -- is performed in the context of a highly controllable ultracold fermionic atom experiment at LENS, which observed the transition from coherently oscillating to decaying supercurrents~\cite{PhysRevLett.120.025302}. Our time-dependent analysis
unambiguously identifies the distinct microscopic origins of emerging dissipative dynamics across the weakly- (BCS) and strongly- (UFG) interacting limits, despite global system quantities exhibiting similar features.

\begin{figure*}[t]
\begin{center}
\includegraphics[width= \textwidth]{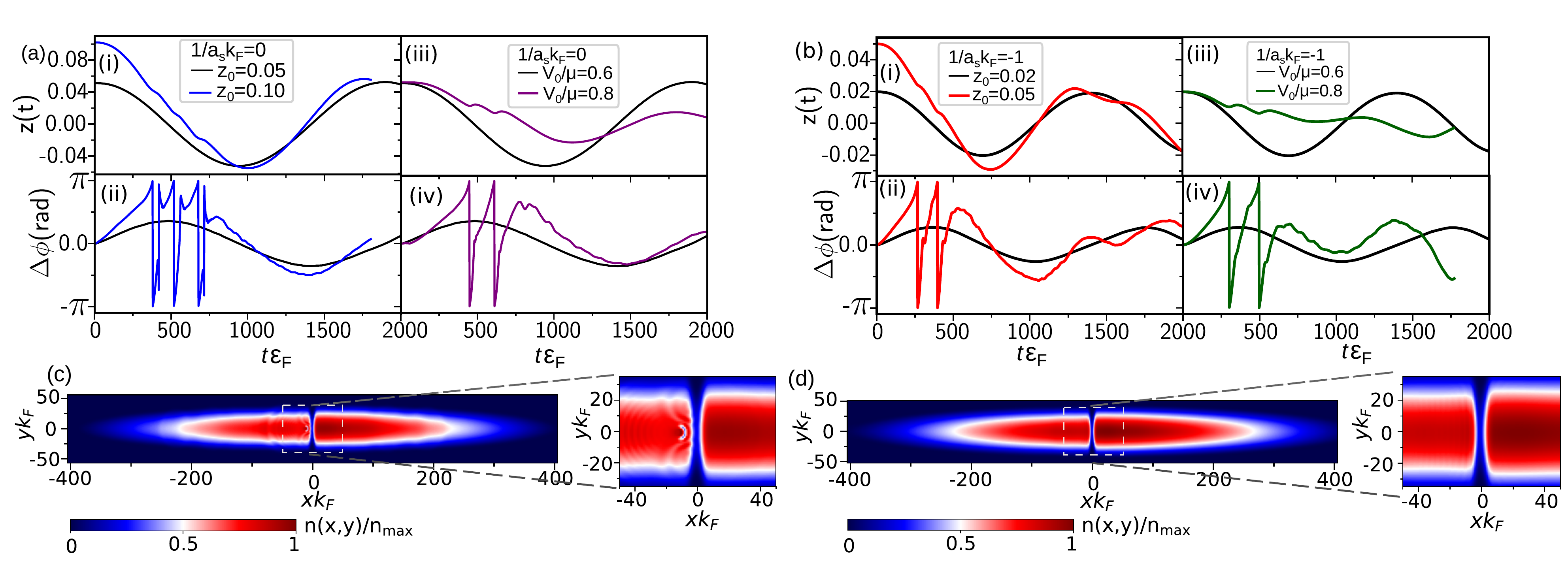}
\vspace{-0.6cm}
\caption{Dynamical regimes for a fermionic Josephson junction:
The time evolution of population imbalance $z(t)$ [(i),(iii)] and relative phase $\Delta \phi (t)$ [(ii),(iv)] in  UFG (a)  and BCS limit (b). The system can be driven into the dissipative regime either by increasing the value of the initial imbalance $z_0$ at fixed barrier height $V_0/\mu=0.6$ (i)-(ii) or by increasing the barrier strength while keeping the initial imbalance unchanged (iii)-(iv).
Snapshots of density distributions $n(x,y)$ (scaled to its maximum value $n_{\mathrm max}$) at a time close to the phase slippage event (abrupt change of $\Delta \phi$ by $2\pi$) for UFG (c) and BCS (d) respectively. In these simulations we used  $z_0=0.15$ and $V_0/\mu=0.6$. The white dashed rectangular box indicates the barrier region with the zoomed-in plot on the right side.
}
\label{ref:Fig1}
\vspace{-0.4cm}
\end{center}
\end{figure*}
\ssec{Theoretical model.}
The BCS regime of the superfluid Fermi gas is studied by means of the time-dependent
Bogoliubov-de Gennes (BdG) equations~\cite{WilhelmZwerger2012}; Up to date, essentially all such fermionic Josephson junction studies were limited to considerations of either static cases ~\cite{PhysRevLett.99.040401,Spuntarelli2010,PhysRevB.102.144517}, or 1D scenarios~\cite{zou2014josephson} (with a simplified two-mode model considered in~\cite{Li2020,Li2021}).
The time-dependent BdG equations describe the evolution of quasiparticle wave functions
$\varphi_n({\bf r},t) = [u_n({\bf r},t), v_n({\bf r},t)]^T$, and the Pauli principle is manifested by orthonormality $\int \varphi_n^{\dagger}({\bf r},t)\varphi_m({\bf r},t)d{\bf r}=\delta_{nm}$ that must be satisfied at each time.
The equations read
(using units of $m=\hbar=1$)
\begin{equation}\label{eq:tdbdg}
i\dfrac{\partial}{\partial t} \begin{pmatrix} u_n({\bf r},t)\\ v_n({\bf r},t)\\ \end{pmatrix} =
\begin{pmatrix}h({\bf r},t) & \Delta({\bf r},t)\\ \Delta^*({\bf r},t) & -h^{*}({\bf r},t)\\ \end{pmatrix} \begin{pmatrix} u_n({\bf r},t)\\ v_n({\bf r},t)\\ \end{pmatrix},
\end{equation}
where $h$ is the single quasiparticle Hamiltonian $h({\bf r},t) = -\frac{\nabla^2}{2} + U({\bf r},t) +V_{\rm ext}({\bf r},t)-\mu$ shifted by the chemical potential $\mu$. $U$ is the mean-field potential, which is routinely neglected in the BCS regime, $V_{\rm ext}({\bf r},t)$ is an external potential. The pairing potential $\Delta({\bf r},t)=-g\nu({\bf r},t)$ models Cooper pairing phenomena, where $g = 4\pi a_s$ is the coupling constant defined through the scattering length $a_s$, and the anomalous density reads $\nu=\sum_{E_n>0}v_n^*u_n$, where $E_n$ is quasiparticle energy corresponding to $n$-th quasi-orbital.
In practice, a renormalized coupling constant is used, with the sum evaluated up to a cut-off energy $E_c$ in order to cure the ultraviolet divergence~\cite{Bulgac2002,PhysRevA.76.040502}.
The equations are valid for $|a_s \kF| \lesssim 1$.
The Fermi wave-vector and associated Fermi energy are defined through relation to gas density $n$ as  $\kF=\sqrt{2\eF} = (3\pi^2n)^{1/3}$.
In order to initialize the time-dependent simulation we need to provide the solution of the static variant of Eq.~(\ref{eq:tdbdg}), i.e.~we carry out the replacement $i\frac{\partial}{\partial t}\rightarrow E_n$.
Our analysis is based on the following observables (omitting here, for brevity, position and time dependencies): (i) the particle number density $n=\sum_{E_n>0}|v_n|^2$; (ii) the current $\bm{j}=2\sum_{E_n>0}\textrm{Im}\left[v_n\bm{\nabla} v_n^* \right]$; and (iii)
the pairing field (order parameter) $\Delta$.

The UFG limit is instead modelled within the framework of the
Superfluid Local Density Approximation (SLDA)~\cite{PhysRevA.76.040502, Bulgac2008,Bulgac2012}. Exploiting the concept of Density Functional Theory (DFT), this is a well-tested extension of the BdG scheme beyond the BCS regime,
allowing to grasp the limit of
strong interaction regime with $a_s \kF \to \infty$.
The scale invariance property of the UFG imposes that $\Delta^{(\textrm{UFG})}=-\frac{\gamma}{n^{1/3}} \nu$, and $U^{(\textrm{UFG})}=\frac{\beta(3\pi^2 n)^{2/3}}{2}-\frac{|\Delta|^2}{3\gamma n^{2/3}}$, which is no longer negligible.  The coupling constants $\beta$ and $\gamma$ are adjusted to ensure correct reproduction of basic properties of the uniform gas, like the Bertsch parameter $\xi_0\approx0.4$ and the energy gap $\Delta/\eF\approx 0.5$. Over the years, the SLDA approach has been successfully applied to a variety of problems, like dynamics of topological defects~\cite{Bulgac2011,PhysRevLett.108.150401,PhysRevLett.112.025301,Wlazlowski2018,PhysRevA.103.L051302}, Higgs modes~\cite{PhysRevLett.102.085302}, properties of spin-imbalanced systems~\cite{PhysRevLett.97.020402,Bulgac2008,PhysRevA.104.053322,PhysRevA.100.033613,PhysRevA.104.033304} and even quantum turbulence~\cite{PhysRevA.91.031602,PhysRevA.105.013304}.

\ssec{Parameter Regime and Numerical Implementation.}
Our studies are motivated by the LENS ${}^6\textrm{Li}$ experimental setup presented in Ref.~\cite{PhysRevLett.120.025302}.
However, as the direct numerical solution of the full 3D equations of motion~(\ref{eq:tdbdg}) is beyond reach of present computing systems, we simplify the computation process by adopting an effectively two-dimensional geometry:
in effect, we assume quasi-particle wave functions
of a generic form $\varphi_n({\bf r},t)\equiv\varphi_n(x,y,t)e^{i k_z z}$, and approximate the full 3D harmonic trap potential $V_{\rm ho}({\bf r})\rightarrow V_{\rm ho}(x,y)= m \omega_x^2 (x^2 + \lambda^2 y^2) / 2$, with $\lambda = \omega_y / \omega_x=148/15$ being the aspect ratio, as in the experiment.
We solve the problem on a computational grid of size $N_x\times N_y\times N_z=768 \times 96 \times 24$ (lattice spacing is $dx=1$ and defines the numerical unit of length). The number of
atoms per $z$-plane is $\Ntot/N_z\approx 830$. Simultaneously, we adjust $\omega_x$ such that $\kF\approx 1.1$ in the trap center to ensure the BCS coherence length $\xi=\kF/\pi\Delta \gtrsim dx$, so that topological defects, like quantum vortices, can be numerically resolved.
The double-well geometry of the Josephson junction is engineered by adding  a Gaussian barrier $V_{\rm b}({\bf r}) = V_0 e^{-2x^2/w^2}$ along the x-axis.
The initial configuration corresponds to a slight density imbalance between the two halves of the cloud, which we achieve by adding a slight linear tilt to the potential $V_{\rm t}({\bf r})=\alpha x$, when generating a stationary solution. Then, we remove the tilt and allow the system to follow its dynamics. The number of evolved quasiparticle states is $n\approx 5\cdot 10^5$, which results in solving about one million of coupled and nonlinear PDEs, which we treat by High-Performance Computing techniques.

We investigate the dynamics of Josephson atomic junction for two interaction regimes: the BCS regime with $1/a_s\kF\simeq -1$ (which extends deeper into this regime than recent experiments~\cite{PhysRevLett.120.025302}), and the unitary limit  $1/a_s\kF \simeq 0$. Guided by previous works~\cite{PhysRevLett.124.045301,Xhani2020,KXfiniteT} in the BEC regime ($1/a_s\kF> 1$)
we consider relatively narrow barriers of
width $wk_F=5.2$ (which corresponds to a value of $w/\xi=4$ in the UFG and $1.6$ in the BCS limit) and values of $V_0/\mu < 1$ (which, in the BEC limit, were found to maximize vortex detection likelihood). Within such parameter space, we study  system dynamics as a function of two control parameters: (i) the barrier height scaled to the (mean) chemical potential~$V_0/\mu$, and (ii) the initial population imbalance $z_0\equiv z(t=0)$, where  $z(t)=(N_R(t)-N_L(t))/\Ntot$ with $N_L\,(N_R)$ the  number of atoms in the left (right) reservoir. The visualisation of the numerical setup is presented in Fig.~\ref{ref:Fig1}(c)-(d). We have also checked that conclusions are unchanged if we compare results between regimes for fixed $w/\xi=4$.
\begin{figure}[t]
\begin{center}
\includegraphics[width=1.0\columnwidth]{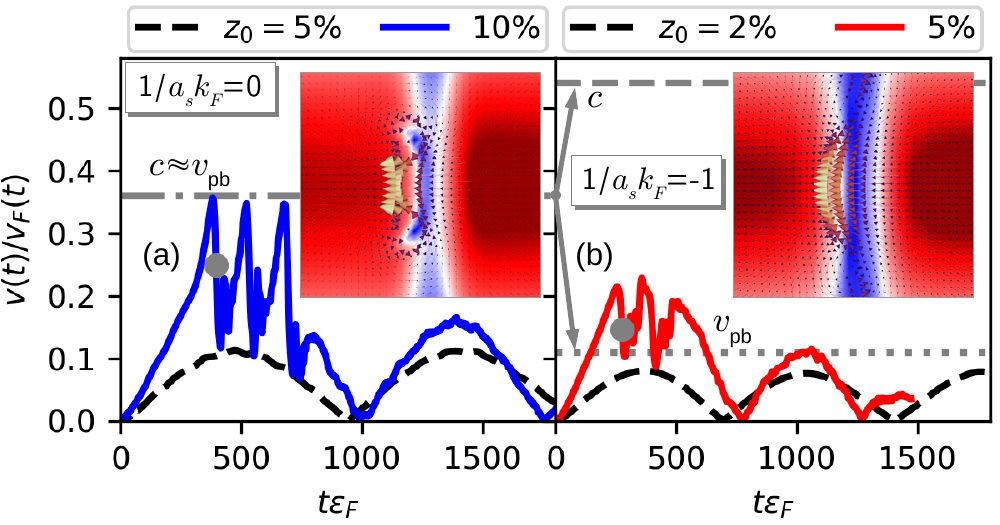}
\vspace{-0.6cm}
\caption{
The time evolution of the flow velocity in the barrier, scaled to the local Fermi velocity $v(t)/v_F(t)$ in the UFG (a) and the BCS limits (b). Dashed (black) lines demonstrate case of coherent oscillations, while solid (blue/red) lines dissipative dynamics.
The grey dotted horizontal lines indicate  the speed of sound  $c$ and the pair-breaking velocity $v_{\textrm{pb}}$. These data are for fixed $w\kF=5.2$ and fixed $V_0/\mu=0.6$.
Insets show velocity fields $v(t)$ topology (arrows) in the barrier regions for the dissipative cases and time moments indicated by (grey) dots. The color map visualises $|\Delta(x,y)|$. The arrows indicate separation of the scales $c$ and $\vpb$ in the BCS regime.
}
\label{ref:Fig2}
\vspace{-0.4cm}
\end{center}
\end{figure}

\ssec{Dynamical Regimes in a Fermi superfluid.} Investigation of the dynamical regimes is performed by studying the temporal profiles of the relative imbalance $z(t)$ and its canonically conjugated variable $\Delta \phi (t)\equiv \phi _L(t)-\phi _R(t)$ [see Fig.~\ref{ref:Fig1}(a)-(b)], where $\phi _{L(R)}$ is the phase of the order parameter extracted at a point in the left (right) side of the barrier. Note that the applied frameworks conserve the total energy, $E_{\textrm{tot}}$~\cite{SM}. Here, we consider transfer of energy stored in the junction to other degrees of freedom. The energy stored in the junction is quantified by the sum of two terms: $[E_C\Ntot^2/8]z^2(t)$ and $E_J[1-\cos\Delta \phi (t)]$, where $E_{J(C)}$ is the Josephson (capacitive) energy~\cite{ZAPATA1998}.
The non-dissipative or coherent regime is characterized by sinusoidal oscillations with constant amplitude of both the variables $z(t)$ and $\Delta \phi (t)$~\cite{Smerzi03,PhysRevLett.79.4950,Xhani2020,PhysRevLett.124.045301,Meier2001}, while dissipative dynamics is reflected by the decaying amplitude for $z(t)$. As expected, the coherent oscillations are observed both in the BCS and UFG regimes for sufficiently low $z_0$ or $V_0/\mu$ [black lines in Fig.~\ref{ref:Fig1}(a)-(b)].

\begin{figure*}[t]
\begin{center}
\includegraphics[width=\textwidth]{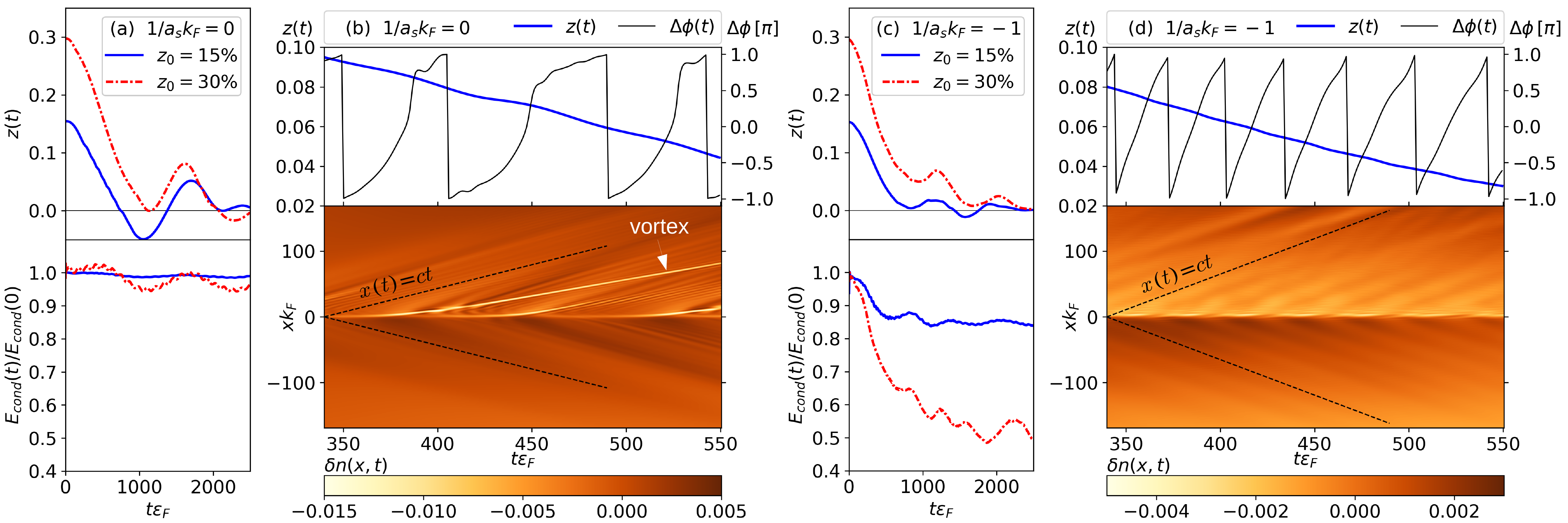}
\vspace{-0.6cm}
\caption{
The time evolution of the relative imbalance $z(t)$ (top) and the condensation energy $E_{\textrm{cond}}(t)$ (bottom) for the unitary (a) and BCS (c) limits.  The data are taken in the dissipative regimes for $V_0/\mu=0.6$ and $w\kF=5.2$ and for two different values of the initial imbalance $z_0$. For the case $z_0=15\%$ we show dynamics within selected time interval of the relative imbalance $z(t)$  and the relative phase difference $\Delta\phi(t)$ (top) together with the carpet $\delta n(x,t)\equiv n(x,0,t)-n(x,0,0)$ (bottom) for the UFG (b) and BCS (d) regimes. Dashed lines in the carpet plot indicate trajectory for objects moving with speed of sound $x=\pm c t$. In the UFG the carpet plot reveals directly both the vortex dipoles and the sound waves.
}
\label{ref:Fig3}
\vspace{-0.4cm}
\end{center}
\end{figure*}

At a critical value of the barrier height or the initial imbalance the system is expected to transition to a dissipative regime \cite{PhysRevLett.120.025302, Valtolina2015,Xhani2020, PhysRevLett.124.045301},
characterized by a damping of the population imbalance oscillation amplitude and by the relative phase showing $2\pi$ jumps, called phase slips.
In the BEC limit, such dissipative dynamics have been interpreted in terms of the generation of vortices and sound waves \cite{PhysRevLett.124.045301, Xhani2020}.
A characteristic of such dissipative dynamics is the presence of kinks in $z(t)$, due to the
backflow caused by the generation of quantum vortices.
Similar dissipative features, with corresponding $2 \pi$ phase jumps in the relative phase are evidently present in the population dynamics in Fig.~\ref{ref:Fig1}(a)-(b) in both the UFG and BCS limits.
Despite very similar features in the above observables, a fundamental difference becomes apparent when considering the corresponding density distribution snapshot in the broader barrier region [Fig.~\ref{ref:Fig1}(c)-(d)] close to the phase slippage moment.
While both vortex pairs and sound waves are found to propagate into the left reservoir in the UFG limit (being more visible for low values of $V_0/\mu$), the corresponding BCS limit instead displays only barely visible (low amplitude) sound wave propagation and no discernible evidence of vortex generation [see also subsequent Fig.~\ref{ref:Fig3}], a feature which is consistent across all simulations executed in the BCS regime.
This points towards potentially different origin and underlying mechanisms of dissipation.
Another generic feature that we find is that as we sweep the interaction from the UFG to BCS limit, for fixed barrier height $V_0/\mu$,
the critical imbalance delimiting the coherent from the dissipative regime becomes smaller.
Moreover, analysis of the current-phase relation at the respective critical imbalances
(at $V_0/\mu=0.8$), reveals
a lower BCS density current compared to UFG regime, consistent with the observed critical current suppression~\cite{PhysRevA.100.063601,Kwon2020,Valtolina2015,PhysRevLett.120.025302}.

\ssec{Dissipative mechanisms.}
It is well-known that  matter flow can be dissipationless in the presence of an obstacle (e.g.~in the form of a barrier), provided its speed does not exceed a critical value. Specifically, the Landau criterion states that if quasiparticle energy in the reference frame of the obstacle,
$\varepsilon(\bm{p})+\bm{p}\cdot \bm{v}_{s}$,
becomes negative, then the excitations carried by this quasiparticle can be created spontaneously. Here by $\varepsilon(\bm{p})$ we denote the quasiparticle energy of momentum $\bm{p}$  in the reference frame where the superfluid is at rest and $\bm{v}_{s}$
is the speed of the flow. The most common type of low-energy excitations are phonons where  $\varepsilon(p)=c p$ and $c$ is the speed of sound.
The phonons can be emitted spontaneously when $v_s>c$.
The speed of sound is related to the equation of state $c^2=\left.\frac{\partial P}{\partial n}\right|_S=\frac{V^2}{N}\left.\frac{\partial^2 E}{\partial V^2}\right|_S$. For the UFG where $E=\xi_0\Effg=\xi_0\frac{3}{5}\Ntot\eF$,
we have $c_{\textrm{UFG}}=\sqrt{\frac{\xi_0}{3}}\vF$, where $\vF$ is the Fermi velocity.
Using BCS theory where $E=\Effg-\frac{3N\Delta^2}{8\eF}$ with $\Delta/\eF=\frac{8}{e^2}e^{\pi/2a_s\kF}$ we obtain
\begin{equation}
c_{\textrm{BCS}}=\sqrt{\frac{1}{3}-\frac{12}{e^4}e^{\pi/a_s\kF}\left[\left( \frac{\pi}{3a_s\kF}\right)^2-\frac{2\pi}{3a_s\kF}+\frac{10}{9}\right]}\vF.
\end{equation}
Qualitatively, the speed of sound increases as we quench the interaction towards the deep BCS regime.

Beside creating phonons, for Fermi systems we may break Cooper pairs and induce quasiparticle excitations with energy $\varepsilon(p)=\sqrt{\left(\frac{p^2}{2}-\mu\right)^2 + \Delta^2}$. Then the Landau criterion leads to a distinct critical velocity, associated with {\it pair breaking},
$\vpb=\sqrt{\sqrt{\mu^2+\Delta^2}-\mu}$.
For the UFG, where $\mu/\eF=\xi_0$ and $\Delta/\eF\approx 0.5$ one finds $ \vpb\approx c_{\textrm{UFG}}\approx 0.36\vF$~\cite{inguscio2007ultracold,WilhelmZwerger2012}, while in the BCS regime where $\Delta$ is exponentially small and $\mu\approx\eF$ we obtain $\vpb\approx\frac{\Delta}{2\eF}\vF\ll c_{\textrm{BCS}}$. Thus, we expect that the pair breaking mechanism will be dominant in the BCS regime. The critical velocities were studied in works~\cite{PhysRevLett.99.040401,PhysRevA.74.042717,PhysRevLett.99.070402,PhysRevLett.114.095301,PhysRevLett.121.225301,PhysRevA.100.033611}.

In Fig.~\ref{ref:Fig2} we present the flow velocity inside the barrier, $v(t)\equiv |\bm{v}(0,t)|$, normalized to the local value of the Fermi velocity, $\vF(t)=[3\pi^2 n(0,t)]^{1/3}$, and compare it to the characteristic scales. The velocity field is extracted via $\bm{v}(\bm{r},t)=\bm{j}(\bm{r},t)/n(\bm{r},t)$.
If the flow remains below both the
speed of sound and the pair breaking velocity, the superfluid continues its motion without dissipation, oscillating coherently between the two reservoirs. However, the dynamics changes if the flow reaches one of the critical values. Consider maximum value of the flow  $v_{\mathrm{max}}\equiv \max[v(t)]$.
In the strongly interacting limit [Fig.~\ref{ref:Fig2}(a)], we find that the maximal detected value is approximately equal to the speed of sound $\vmax\approx c$. Whenever the local speed approaches it, quantum vorticity is nucleated (here in the form of a vortex-antivortex pair), and the flow is reduced abruptly.
Contrary, in the BCS limit, the maximal detected speed is  much lower than $c$, but simultaneously larger than the pair-breaking velocity $v_{\mathrm{pb}}$ calculated previously, i.e.~$v_{\mathrm {pb}}\lesssim \vmax < c $. While transient configurations where the velocity field exhibits swirling patterns inside the barrier are found [inset to Fig.~\ref{ref:Fig2}(b)], nonetheless the phase of the order parameter does not exhibit the expected topology. In other words,  throughout our simulations in the BCS limit, even though the relative phase shows $2\pi$ jumps, we do not  unambiguously detect winding of the phase by $2\pi$ in regions where the velocity field swirls.

The change of the dissipative mechanism becomes more evident in the case of initial imbalances $z_0$, much higher than the critical value. Fig.~\ref{ref:Fig3} demonstrates the system dynamics for $z_0=15\%$ and $30\%$, while keeping other parameters as before. We observe a fast drop of $z(t)$, which starts to oscillate (irregularly) around $z=0$. In the case of the BCS limit [Fig.~\ref{ref:Fig3}(c)], we find that the amplitude of the residual oscillations is much smaller than in the UFG case [Fig.~\ref{ref:Fig3}(a)], suggesting that more dissipation is present in the former case, see~\cite{SM} for more details.
As before, for strong interactions, we observe that quantum vortices and sound waves take away the energy to the bulk (typically the sound wave is generated due to vortex pair annihilation or during its propagation in a density gradient \cite{Xhani2020,PhysRevLett.124.045301}).
Contrary to that, in the weakly-interacting case, only relatively small
amplitude sound waves are observed. Such picture is best vizualized in the renormalized density carpet plots (b,d), in which the color represents the instantaneous density value along the $x$-axis after subtracting the initial value $\delta n(x,t)\equiv n(x,0,t)-n(x,0,0)$.

In order to quantify the importance of the pair-breaking mechanism, we calculate the change in the condensation energy. According to the BCS theory, the appearance of a Cooper-pair condensate lowers the energy of the (uniform) system by $\frac{3\Delta^2}{8\eF}\Ntot$. Using the local density approximation we define the condensation energy for the non-uniform system as $E_{\mathrm{cond}}=\int \frac{3}{8} \frac{|\Delta (\mathbf{r})|^2}{\eF (\mathbf{r})} n(\mathbf{r}) d\mathbf{r}$. The change of $E_{\mathrm{cond}}$ is shown in the bottom panels (a) and (c) of Fig.~\ref{ref:Fig3}. The difference between the UFG and the BCS regimes is now evident. For the unitary gas, the condensation energy can, to good approximation, be regarded as a conserved quantity during the dynamics. Only for the most extreme case studied by us, $z_0=30\%$, we find that it drops by a few percent (over our probed timescale $t \varepsilon_F$). On the other hand, for the BCS gas, the energy stored in the condensate decreases noticeably in time. For example, in the analogous case $z_0=30\%$, we observe a drop of $E_{\mathrm{cond}}$ by about half, a striking manifestation of the depletion of the Cooper-pair condensate.
It has to be noted that although the results presented above indicate the main mechanisms of energy dissipation,
the accurate determination of the dissipation rate would require longer trajectories to be able to extract irreversible energy transfer.

\ssec{Conclusions.}
The change in the underlying physical mechanism giving rise to dissipative dynamics in Fermi
superfluids from vortex nucleation to Cooper-pair breaking can be deduced based on simple arguments related to
the ordering of the velocity scales $c$ and $\vpb$ . However, it does not provide information on how this change will
be manifested in the real-time (population) dynamics. Surprisingly, global characteristics like imbalance or the
phase difference, which are used as primary probes in experiments, display similar patterns irrespectively of the
operating mechanism. Their time-dependence is similar to the experimental findings~\cite{Valtolina2015,PhysRevLett.120.025302}.
At unitarity, the main dissipative mechanism is related to the phase-slippage, caused by emission and propagation of quantum vortices, and associated sound waves, as observed experimentally (through a barrier removal protocol which enhances vortex lifetime).
Probing deeper in the BCS regime ($1/a_s \kF\simeq -1$) than is presently accessible in experiments ($1/a_s \kF\simeq -0.6$),
we go beyond indirect experimental measures, such as the observed critical current suppression~\cite{PhysRevLett.120.025302}, to
 quantify pair breaking in terms of a decaying Cooper-pair condensation energy, finding its role to be enhanced with increasing population imbalance beyond the critical value, and to dominate the picture without any discernible direct role of vortex dynamics in this regime.
In both cases, the emitted energy is ultimately converted into heat.
Our work provides a deeper understanding of dissipation mechanisms in ultracold fermionic superfluids across the BCS-BEC crossover.

\begin{acknowledgments}
The calculations were executed by means of the W-SLDA Toolkit~\cite{WSLDAToolkit}. Detailed instructions allowing for reproduction of the presented results are given in the Supplementary Material~\cite{SM}.
We thank G. Roati, W. J. Kwon and M. M. Forbes for fruitful discussions.
Calculations were executed by GW and MT, data analysis was performed by GW, KX and MT. All authors contributed to research planning, interpretation of the results and manuscript writing.
We acknowledge PRACE for awarding us access to resource Piz Daint based in Switzerland at Swiss National Supercomputing Centre (CSCS), decision No. 2021240031.
This work was supported by the Polish National Science Center (NCN) under Contracts No. UMO-2017/26/E/ST3/00428 (GW), UMO-2019/35/D/ST2/00201 (MT) UMO-2017/27/B/ST2/02792 (PM), and and  EU’s
Horizon 2020 research and innovation programme under the Qombs project FET Flagship on Quantum Technologies GA no. 820419 (KX).
\end{acknowledgments}

\bibliography{FJJ}

\begin{center}
{\bf Supplementary Material for:}\\
{\bf ``Dissipation Mechanisms in Fermionic Josephson Junction''}\\
\end{center}
\setcounter{figure}{3}

{\small
In this Supplementary Material we demonstrate the accuracy to which the total energy is conserved during time evolution and provide a comparison of obtained dissipation rates between BCS and UFG regimes. We also list the reproducibility packs which are attached and provide the complete information needed to restore the results from the
main paper. }

\section{Energy conservation quality}
The applied formalism of BdG and SLDA conserve total energy:
\begin{equation}
    E_\textrm{tot}(t)=\int \mathcal{E}(\bm{r},t)\,d^3\bm{r} + \int V_{\textrm{ext}}(\bm{r},t)n(\bm{r},t)\,d^3\bm{r},
\end{equation}
where $\mathcal{E}$ is the energy density. For BdG it reads (for brevity we skip position and time dependence of densities, and we use units $m=\hbar=dx=1$):
\begin{equation}
    \mathcal{E}_{\textrm{BdG}}=\frac{\tau}{2}+g|\nu|^2,
\end{equation}
and for SLDA:
\begin{equation}
    \mathcal{E}_{\textrm{SLDA}}=\frac{\tau}{2}+\beta\frac{3(3\pi^2)^{2/3}n^{5/3}}{10}+\gamma\frac{|\nu|^2}{n^{1/3}}.
\end{equation}
The functional is defined through densities: normal $n(\bm{r},t)$ and anomalous $\nu(\bm{r},t)$ as defined in the main text, and kinetic density
$\tau(\bm{r},t)=\sum_{E_n>0}|\nabla v_n(\bm{r},t)|^2$.
Comprehensive discussion of the functionals is given in Ref.~[28].
The dissipation process considered here transfers energy stored in the Josephson junction oscillations to other degrees of freedom.

The associated equations of motions emerge as a result of the stationarity condition of the action:
\begin{equation}
S=\int_{t_0}^{t_1} \left( \langle 0(t)| i\frac{d}{dt}| 0(t) \rangle - E_{\textrm{tot}}(t) \right) dt,
\end{equation}
where $|0(t) \rangle$ denotes the quasiparticle vacuum at time $t$.
The equations have the form given by Eq.~(1) in the main text, with
\begin{align}
    h({\bf r},t) &= -\frac{\nabla^2}{2} + \frac{\delta\mathcal{E}}{\delta n} +V_{\rm ext}({\bf r},t),\\
    \Delta({\bf r},t) &= -\frac{\delta\mathcal{E}}{\delta \nu^*}.
\end{align}

\begin{figure}[th]
\begin{center}
\includegraphics[width=\columnwidth]{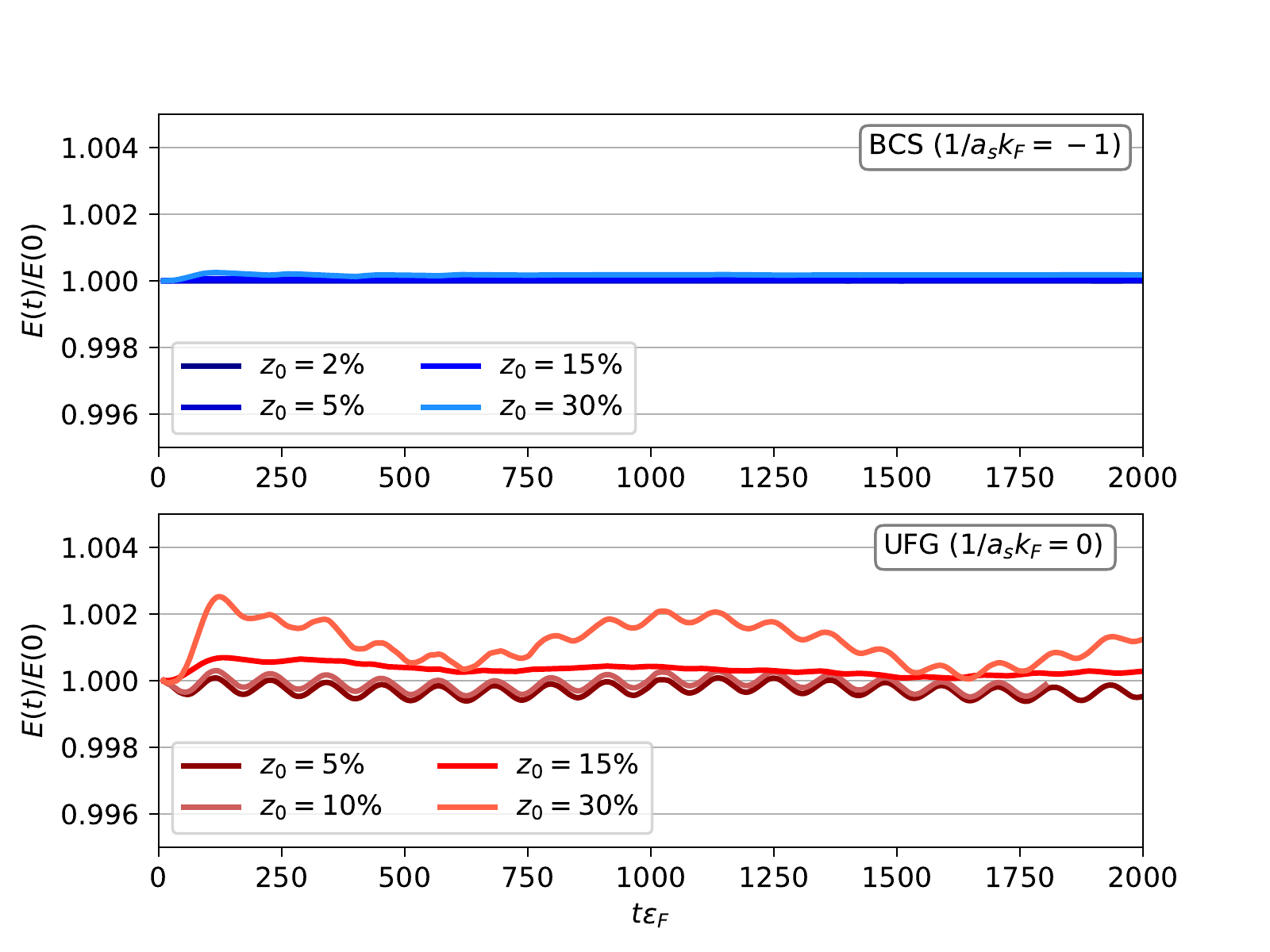}
\caption{Total energy as a function of time for BCS (top) and UFG (bottom) regimes. Lines corresponds to various initial population imbalances. In all cases the total energy is conserved. Visible fluctuations for unitary Fermi gas are due to the regularization procedure. In the worst case (UFG, $z_0=30\%$) the relative departure from the mean value does not exceed $0.13\%$.
}
\label{fig:E_conservation}
\end{center}
\end{figure}

In Fig.~\ref{fig:E_conservation} we present the quality of the total energy conservation for simulations presented in the main paper.
In the numerical realization the total energy is conserved with a high accuracy. In the case of the SLDA simulations for unitary Fermi gas we observe small energy fluctuations around the constant values. In the worst case, the relative size of the fluctuations does not exceed $0.13\%$. The fluctuations are due to the regularization procedure. Namely, the total energy is finite, but contributions taken separately from terms proportional to kinetic energy density $\tau$ and anomalous density $\nu$ are divergent. To cure this problem, an energy cut-off $E_c$ is introduced, and the pairing coupling constant ($g$ for BdG and $\gamma$ for SLDA) is accordingly redefined. For our regularization scheme we use the prescription provided in Refs~[25,26]. As we increase the cut-off energy scale $E_c$, the quality of the energy conservation also increases. In the calculations we have used a fixed value of $E_c=\pi^2/2$, which translates into $E_c\approx 8\varepsilon_F$ for calculations in the BCS regime, and
$E_c\approx 7\varepsilon_F$ for calculations in the UFG regime.

\section{Decay rate of the population imbalance}
In order to provide deeper insight into the decay dynamics of the population imbalance $z(t)$ we provide plots showing its derivative $dz/dt$ for data presented in Fig.~3 of the main text. The derivative has been computed using the finite difference formula. We find that the $z(t)$ decays faster toward residual oscillations around zero ($dz/dt$ oscillates close to zero) in the BCS regime, as compared to the UFG case.
In fact, if no dissipation were present, the current would oscillate sinusoidally in time maintaining a constant amplitude. However, under the presence of dissipative mechanisms this does not occur and the current amplitude evidently decays in time in our simulations.
\begin{figure}[h]
\begin{center}
\includegraphics[width= \columnwidth]{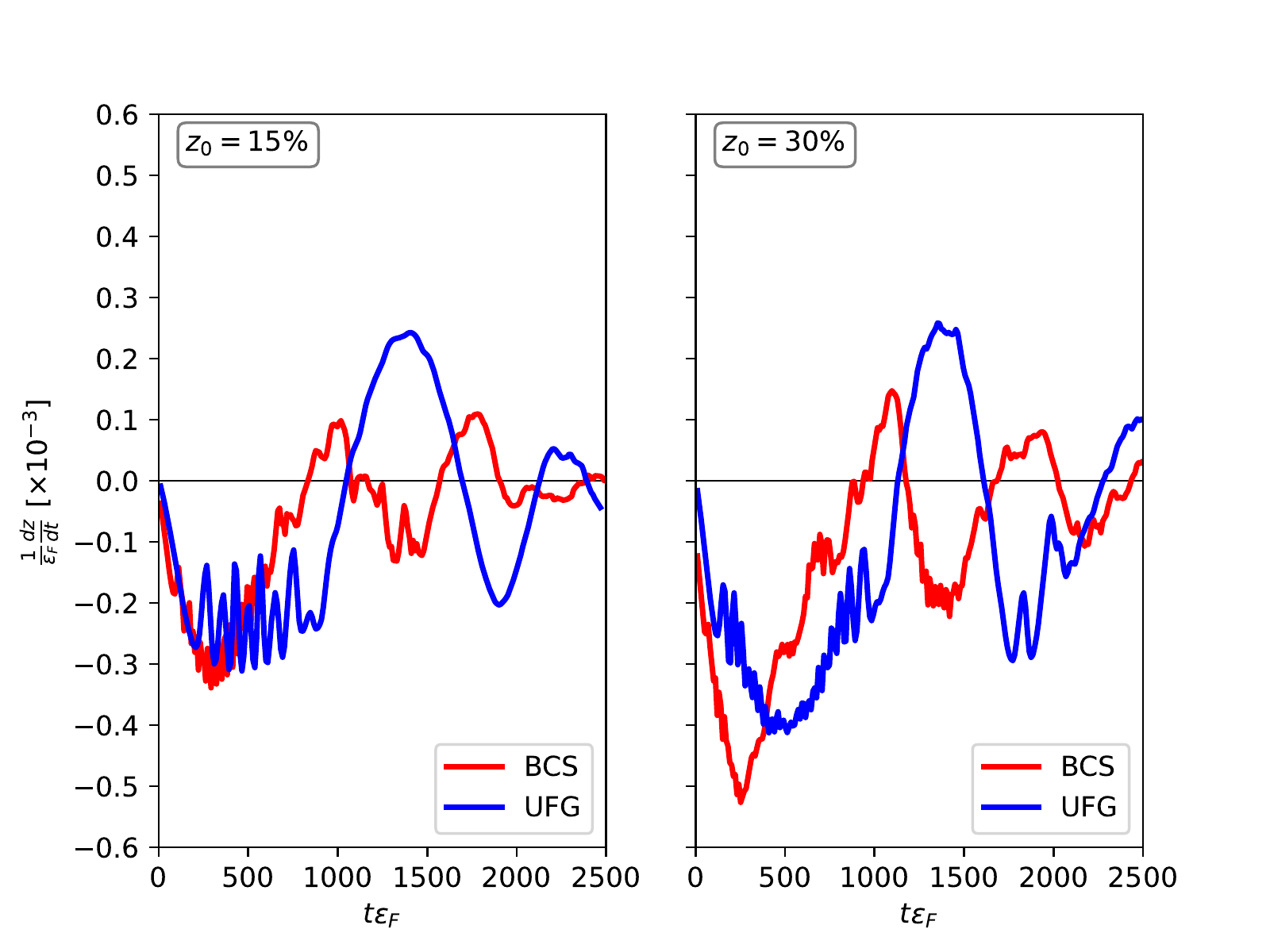}
\caption{Decay rate $dz/dt$ for the data presented in Fig.~3 of the main text. The derivative is computed numerically using the finite difference formula.
}
\label{fig:dzdt}
\end{center}
\end{figure}
\begin{figure}[h]
\begin{center}
\includegraphics[width= \columnwidth]{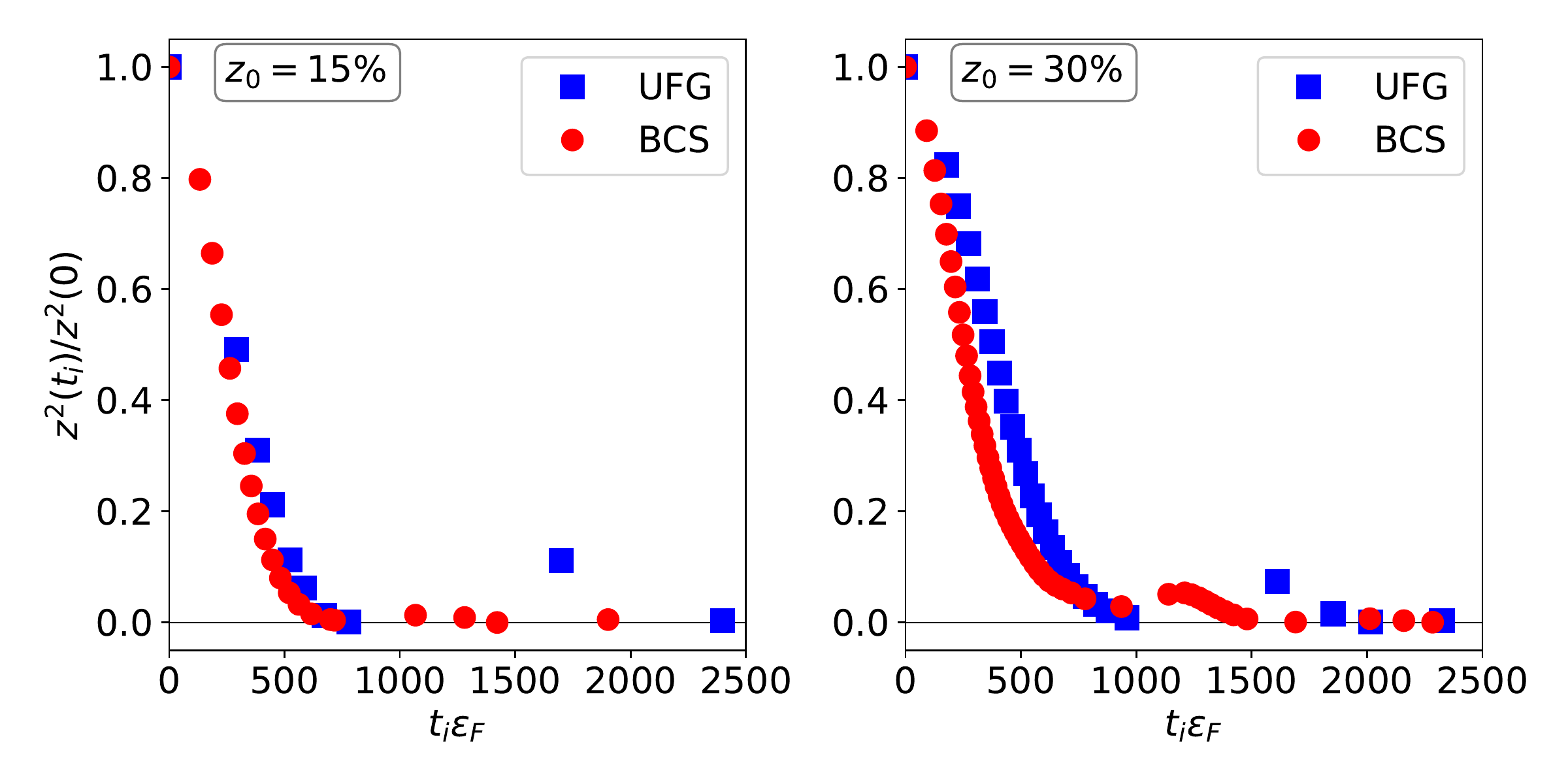}
\caption{The ratio $z^2(t_i)/z^2(0)$ evaluated for time moments $t_i$ where the relative phase between reservoirs vanishes, i.e.~$\Delta \phi (t_i)=0$. The ratio is computed for the data series presented in Fig.~3 of the main text.
}
\label{fig:z2}
\end{center}
\end{figure}


To provide a more robust quantification of the decay rate, we utilize here an effective formula for the energy stored in the Josephson oscillations~[14]:
\begin{equation}
    E_{JJ}(t)=\frac{E_C\Ntot^2}{8}z^2(t)+E_J[1-\cos\Delta \phi (t)],
    \label{eq_en}
\end{equation}
where $E_{J}$ and $E_{C}$ are the Josephson and capacitive energies, respectively. Note, that
the Eq.~(\ref{eq_en}) is valid when no dissipation is present -- otherwise non-coherent or dissipative terms are also anticipated to further contribute to the system energy
as described in Ref.~[42]. These terms are expected to originate from the couplings between the condensate and the noncondensate states. In the BEC regime, the latter consist of phonon-like excitations  at $T=0$ and/or a  thermal component at finite $T$. In the BCS regime, the broken pairs will also contribute to the noncondensate states.
Despite such issues, it is nonetheless rather instructive to focus our analysis specifically on those times  $t_i$ at which $\Delta \phi (t_i)=0$. Then, the energy stored in the Josephson junction can be estimated as $E_{JJ}(t_i)\sim z^2(t_i)$. This quantity is shown in Fig.~\ref{fig:z2}. We clearly observe that the Josephson junction energy is dissipated.
We emphasize that more detailed studies are required to quantify dissipation rates through the vortex nucleation and the pair breaking. In particular this should be done by extracting the appropriate transport coefficient.

\section{Reproducibility packs}
The attached reproducibility packs provide the complete information needed to restore the results from the main paper. Precisely, these are settings that one needs to apply within W-SLDA code to reproduce two selected cases:
\begin{description}
\item [bcs.zip] settings to restore data for BCS regime with $z_0=5\%$ and $V_0/\mu=0.6$ (red line in Fig.~1(b) of the main text).
\item [ufg.zip] settings to restore data for UFG regime with $z_0=5\%$ and $V_0/\mu=0.6$ (black line in Fig.~1(a) of the main text).
\end{description}
The attached \verb|README.md| file contains detailed instruction how to apply these settings to W-SLDA code. Remaining data series can be reproduced in an analogous way, by changing $z_0$ and $V_0/\mu$ parameters.

\end{document}